\documentclass[12pt,prd,draft,nofootinbib]{revtex4}

\usepackage{amssymb}
\usepackage{amsmath}
\usepackage{bm}
\usepackage{latexsym}

\newcommand{\beq}{\begin{eqnarray}}
\newcommand{\eeq}{\end{eqnarray}}
\newcommand{\be}{\begin{equation}}
\newcommand{\ee}{\end{equation}}
\def\la{\mathrel{\mathpalette\fun <}}
\def\fun#1#2{\lower3.6pt\vbox{\baselineskip0pt\lineskip.9pt
\ialign{$\mathsurround=0pt#1\hfil ##\hfil$\crcr#2\crcr\sim\crcr}}}
\newcommand{{\SD}}{\rm SD}

\newcommand{\veJ}{\bm J}
\newcommand{\veL}{\bm L}
\newcommand{\vel}{\bm l}

\newcommand{\ver}{\bm r}
\newcommand{\vesig}{\bm \sigma}

\newcommand{\vep}{\bm p}
\newcommand{\veS}{\bm S}
\newcommand{\ves}{\bm s}

\begin{document}

\title{Higher excitations of the $D$ and $D_s$ mesons}

\author{\firstname{A.M.}~\surname{Badalian}}
\email{badalian@itep.ru} \affiliation{Institute of Theoretical and
Experimental Physics, Moscow, Russia}

\author{\firstname{B.L.G.}~\surname{Bakker}}
\email{b.l.g.bakker@vu.nl} \affiliation{Department of Physics
and Astronomy, Vrije Universiteit, Amsterdam, The Netherlands}

\date{\today}

\begin{abstract}
The masses of higher $D(nL)$ and $D_s(nL)$ excitations are shown
to decrease due to the string contribution, originating from the
rotation of the QCD string itself: it lowers the masses by 45 MeV
for $L=2~(n=1)$ and by 65 MeV for $L=3~(n=1)$. An additional
decrease $\sim 100$~MeV takes place if the current mass of the
light (strange) quark is used in a relativistic model. For
$D_s(1\,{}^3D_3)$ and $D_s(2P_1^H)$ the calculated masses agree
with the experimental values for $D_s(2860)$ and $D_s(3040)$, and
the masses of $D(2\,{}^1S_0)$, $D(2\,{}^3S_1)$, $D(1\,{}^3D_3)$,
and $D(1D_2)$ are in agreement with the new BaBar data. For the
yet undiscovered resonances  we predict the masses
$M(D(2\,{}^3P_2))=2965$~MeV, $M(D(2\,{}^3P_0))=2880$~MeV,
$M(D(1\,{}^3F_4))=3030$~MeV, and $M(D_s(1\,{}^3F_2))=3090$~MeV. We
show that for $L=2,3$ the states with $j_q=l+1/2$ and $j_q=l-1/2$
($J=l$) are almost completely unmixed ($\phi\simeq -1^\circ$),
which implies that the mixing angles $\theta$ between the states
with $S=1$ and $S=0$ ($J=L$) are $\theta\approx 40^\circ$ for
$L=2$ and $\approx 42^\circ$ for $L=3$.

\end{abstract}

\maketitle

\section{Introduction}

Till recently only the low-lying $1S$ and $1P_J$ states of the $D$ and
$D_s$ mesons were known from experiment \cite{ref.1}. The
situation has changed recently owing to discoveries of new $D_s$
resonances: $D_s(2710)$ \cite{ref.2,ref.3}, $D_s(2860)$
\cite{ref.2,ref.4}, and $D_s(3040)$ \cite{ref.4}. Also last
year, new $D(L)$ states were observed by the BaBar Collaboration
\cite{ref.5}: $D(2550)$, $D^*(2600)$, $D_J(2750)$, and $D_J^*(2760)$, and in
\cite{ref.6} the mass of $D(1\,{}^3P_0)$ was measured with a
good accuracy.

The quantum numbers and decay modes of the new resonances were
intensely discussed in a large number of recent studies
\cite{ref.7}--\cite{ref.14} and also before in
\cite{ref.15}--\cite{ref.25}. For $D_{s1}^*(2710)$ the quantum
numbers $J^P=1^-$ were assigned \cite{ref.4,ref.7}, although the
analysis in Ref.~\cite{ref.8} does not exclude that $D_{s1}(2710)$
is an admixture of $D_s(2\,{}^3S_1)$ and $D_s(1\,{}^3D_1)$. The
relatively narrow resonance $D_{sJ}(2860)$ with $\Gamma=48\pm 3$
(stat)~MeV is mostly considered as the $L=2$ state with $J^P=3^-$
\cite{ref.8}--\cite{ref.10}, while in Ref.~\cite{ref.25} the
resonance with close values of the mass and width  has the quantum
numbers $J^P=0^+$. The wide resonance $D_{sJ}(3040)$ is mostly assumed
to be the $2P$ state with $J^P=1^+$ \cite{ref.11}.

The quantum numbers of the new $D^0$ resonances and their isotopic
partners were discussed in \cite{ref.12}--\cite{ref.14}, where the
broad resonance $D(2550)$ ($\Gamma\sim 130$~MeV) is considered as
the singlet $2\,{}^1S_0$ state, for which a large width as in
experiment was obtained in Ref.~\cite{ref.12}, while much smaller
total width was calculated in \cite{ref.13}. The resonance
$D^*(2600)$ with $\Gamma=93\pm 6\pm 13$~MeV is consistent with the
excited $2\,{}^3S_1$ state \cite{ref.14}, or an admixture of the
$2\,{}^3S_1$ and $1\,{}^3D_1$ states with large mixing angle
\cite{ref.12, ref.13}. The resonances $D_J(2760)$ and $D_J(2750)$
have relatively small widths, $\Gamma\sim 60-70$~MeV, and for them
the quantum numbers $J^P=3^-$ and $J^P=2^-$, respectively, were
assigned in Ref.~\cite{ref.13,ref.14}, while in
\cite{ref.19}(the second paper) these two resonances are considered as
the same $1\,{}^3D_1$ state with $J^P=1^-$.

These new data are extremely important for the theory to better
understand the $q\bar Q$ dynamics and test predictions made in a large
variety of models \cite{ref.15}--\cite{ref.25}, some of which were made
long ago \cite{ref.15, ref.16}. For low-lying states, the
theoretical predictions are mostly in agreement with experiment
within $20-50$~MeV accuracy, although the parameters used may
be very different. This is not surprising, because the very
masses of low-lying states are usually used as a fit to determine the
quark masses and parameters of the potentials. On the contrary, for
higher states different predictions for the masses and the
fine-structure (FS) splittings were obtained in different models.

The new experimental data on the hyperfine (HF) splittings show that their
 values, $\simeq 70$~MeV, coincide for $D_s^*(2710)$ and $D_s(2638)$
\cite{ref.26}, $D^*(2610)$ and $D(2540)$. The latter HF splitting was
predicted in Ref.~\cite{ref.27}, if a ``universal" coupling $\alpha_{\rm
HF}=0.31$ is used in the HF potential.

Also the experimental mass differences between $D_s^*(2710)$ and
$D^*(2610)$, $D_s(2638)$ and $D(2540)$, $D_s(2860)$ and $D(2760)$,
appear to be $\simeq 100$~MeV, the only exception being the
$D_s(1P)$ multiplet, where on the one hand, the masses of
$D_{s2}^*(2573)$ and $D_2^*(2460)$, $D_{s1}(2535)$ and $D_1(2422)$
also differ by $\sim 110$~MeV, while
$M(D_{s0}^*(2317)-M(D_0(2300)\approx 25$~MeV and
$M(D_{s1}(2460)-D_1(2430) \approx 30$~MeV are much smaller. Such
small FS splittings cannot be explained within
the ``universal" description of fine structure used in
Refs.~\cite{ref.15, ref.16, ref.18}.

This discrepancy has stimulated a lot of studies to understand why
$D_{s0}^*$ and $D_{s1}$ have such small widths, $\Gamma< 3.8$~MeV
\cite{ref.1}, and large mass shifts. Later it was understood that
the dynamics of the $D_{sJ}(1P)$ multiplet is different for the
states with the total angular momentum of the $s$ quark $j_s=1/2$
and $j_s=3/2$ \cite{ref.21}--\cite{ref.25} and the bispinor
structure of the $D_s(1P)$ wave function (w.f.) and the w.f. of
the $D(1S)$ mesons in the decay channel are very important. Two
factors provide a large hadronic shift: the nearby $S$-wave
threshold and the large overlap integral between the upper
components of the $D_{sJ}(1P)$ w.f. with $j_s=1/2$ and the lower
components of the w.f. of the $D$-meson in the decay channel
\cite{ref.23}.

Surprisingly, there are no large mass shifts for the other excited
$D$ and $D_s$ states observed and, as a whole, the single-channel
description turns out to be a useful tool to understand the
general structure of the $D$ and $D_s$ spectra and FS splittings,
and to predict the masses of the yet undiscovered resonances. Till
now one of the best predictions for the meson masses of the
low-lying states were obtained in the QCD motivated relativistic
quark model (RQM), already in 1985 \cite{ref.15}.

In contrast to the low-lying states, discrepancies show up for
higher states, which may reach $\sim 100$~MeV. For example, for
$D_s(3\,{}^1S_0)$ the masses 3097 MeV from the paper CTLS \cite{ref.7}
and 3259 MeV \cite{ref.16}, and for $M(D_s(2\,{}^3P_2)$ the values
$3041$~MeV \cite{ref.7}, and 3157 MeV \cite{ref.16, ref.17} were
obtained, showing differences $\geq 100$~MeV. The reasons why they
occur will be discussed in the present paper.

A comparison of the results obtained in different models is
simplified, if the same value of the string tension $\sigma$ is
used. The choice of $\sigma$ is of great importance, because the
meson mass is proportional to $\sqrt \sigma$ for the linear
potential $\sigma r$, which dominates for higher states. However,
much different values $\sigma$ are used in potential models: a
large $\sigma\sim 0.26$~GeV$^2$ in \cite{ref.16, ref.17} and small
$\sigma=0.115$~GeV$^2$ in \cite{ref.24}, $\sigma=0.14$~GeV$^2$ in
\cite{ref.20}. Here we use $\sigma=0.18$~GeV$^2$, which follows
from the analysis of the Regge trajectories for light mesons, and
was already used in Refs.~\cite{ref.15, ref.19}. Taking the same
$\sigma$, one can establish common features and differences
between the relativistic string Hamiltonian (RSH) \cite{ref.28}
used here and the RQM developed in
Refs.~\cite{ref.15, ref.19}. In particular, we show that the
choice of the current light (strange) quark mass is of special
importance in relativistic models.

The only uncertainty in our calculations comes from the gluon-exchange (GE)
potential, since at present there is no consensus about the value
of the vector coupling at large distances, called the freezing
constant or the critical constant $\alpha_{\rm crit}$. In
Ref.~\cite{ref.15} the value $\alpha_{\rm crit}=0.60$ was used and
the variation of $\alpha_{\rm crit}$ in the range $0.60\pm 0.10$
produces rather small changes, $\leq 20$~MeV, in the spin-averaged
masses for higher states. However, the value of the strong
coupling in spin-orbit and tensor potentials, $\alpha_{\rm
FS}(\mu)$, is also not fixed now, in contrast to the FS in heavy
quarkonia, where the scale $\mu$ and second order perturbative
corrections are known \cite{ref.29}, giving $\alpha_{\rm FS}(\mu)$
smaller than $\alpha_{\rm crit}$ \cite{ref.30}. In our analysis of
the FS here, we shall test different values of $\alpha_{\rm FS}$ to fit
new experimental data on the $D(1D)$ multiplet.

In our paper we concentrate on the multiplets with $L=2,3$; to
calculate mixing angles for states with $J=L$ we use the basis
${\bm j}_q^2$ from Ref.~\cite{ref.22}, where ${\bm j}_q=\vel+{\bm
s}_q$ is the total angular momentum of the light (strange) quark
and the total spin ${\bm J}={\bm j}_q+{\bm s}_Q$ is the sum of the
light-quark total angular momentum and the spin of the heavy
quark. It appears that for higher states with $J=l$ the states
with $j_q=l+1/2$ and $j_q=l-1/2$ are in fact unmixed,
$|\phi(nl)|\approx 1^\circ$, and this result may be important to
study different decay modes of heavy-light mesons. Owing to the
known relation between the mixing angle $\phi(nl)$ and the mixing
angle $\theta(nL)$ in the $S^2$ scheme (or $\veL\veS$ scheme with
$\veJ=\veL+\veS$), the states with the spin $S=1$ and $S=0$
($J=L$) appear to be mixed with large mixing angle, e.g.
$\theta(1D)=40.2^\circ$.

\section{Relativistic string corrections}

Here we use the RSH, derived for spinless quarks and antiquarks
\cite{ref.28}, while all spin-dependent interactions are
considered as a perturbation. To calculate the spectra of the
heavy-light mesons this approach has some advantages as compared
to the use of the Dirac equation (DE) and considering the heavy
quark contribution as $1/m_Q$ corrections \cite{ref.16, ref.17}.
As shown in Ref.~\cite{ref.31}, for a scalar potential the
solutions of the DE have an important property: the spectrum is
symmetric under the reflection of the eigenvalues (e.v.),
$\epsilon_n\rightarrow -\epsilon_n$, so that negative energy
states are in fact not present in the spectrum of a heavy-light
meson.

Moreover, from the expression for the squared e.v. $\epsilon_n^2$
of the DE (with a given $l=l_q$ and $j=j_q$ -- the total angular
momentum of a light quark) it follows that the mass difference
between neighbouring states is equal to
$\epsilon_{n+1}^2-\epsilon_n^2 \simeq
4\sigma_D+\ln(\frac{\epsilon_{n+1}^2}{\sigma_D|\kappa|}) -
\ln(\frac{\epsilon_n^2}{\sigma_D|\kappa|})$ \cite{ref.31}, where
$\sigma_D$ is the string tension used in the DE and the constant
$\kappa$ enters the Coulomb interaction, -$\frac{\kappa}{r}$. For the
DE this mass difference (for a given $\sigma$) appears to be
significantly smaller than that in the RSH and the RQM, where it
is equal to $4\pi\sigma$.

Just to compensate such a small spacing between radial excitations
the larger value of the string tension, $\sigma_D\simeq
0.26$~GeV$^2$, is needed \cite{ref.16, ref.17} (in both cases the
$1/m_Q$ corrections were taken into account). However, it remains
unclear why in Ref.~\cite{ref.16} the calculated values
$M(D_s(1\,{}^3P_0))=2487$~MeV and $M(D_s(1P_1^{\rm H})) = 2605$~MeV
(for $j_q=1/2$) are similar to the numbers obtained in the
RQM~\cite{ref.15, ref.23}, while much smaller values, $2325$~MeV
and $2467$~MeV, were calculated within a similar approach in
Ref.~\cite{ref.17}. Here we will mostly compare our results with
those models \cite{ref.15, ref.19}, where the same
$\sigma=0.18$~GeV$^2$ was used, and draw definite conclusions
about the dynamics of the $q\bar Q$ interaction.

The RSH $H=H_0+H_{\rm str}$  for spinless quarks and antiquarks was
derived in instantaneous approximation \cite{ref.28} and has the
following characteristic features:

\begin{enumerate}

\item The QCD string, besides a standard rotation of a quark and an
antiquark, rotates itself, giving an additional contribution to a
Hamiltonian, $H_{\rm str}$. For heavy-light mesons such string
corrections are not large, $\sim 30-70$~MeV (for $L=1,2,3)$, and
can be considered as a perturbation \cite{ref.18}, while in light
mesons the string corrections may dominate for states with large
$L$ \cite{ref.32}.

\item In the unperturbed Hamiltonian $H_0=T + V_{\rm B}$ the
kinetic term $\rm T$ \cite{ref.28} is
\begin{equation}
 T=\frac{\omega_1}{2}+\frac{m_1^2}{2\omega_1}+\frac{\omega_2}{2}+
 \frac{m_2^2}{2\omega_2}+\frac{\vep^2}{2\omega_{\rm red}},
\label{eq.01}
\end{equation}
where by derivation the quark mass cannot be chosen arbitrarily and
must be equal to the current mass $\bar m_q$ for the $u,d$, and $s$
quarks and the pole mass $m_Q$ for a heavy quark, thus taking into
account perturbative corrections to the heavy quark mass. In our
calculations $\bar m_q=0$ for the $u,d$ quarks, $m_s\simeq \bar
m_s(1~{\rm GeV})=200$~MeV for the $s$ quark, and the conventional
pole mass $m_c(\rm pole)=1.42$~GeV for the $c$ quark \cite{ref.1} is
used. This choice of $m_s$ is similar to that in Ref.~\cite{ref.16},
where $m_s=220$~MeV was used in the DE, being larger compared to $\bar
m_s(2~{\rm GeV})=95\pm 20$~MeV at the scale $\mu=2$~GeV \cite{ref.1}. The
reason for that difference possibly originates from the fact that in
the Hamiltonian approach the $s$-quark current mass $\bar m_s(\mu)$
enters at a smaller scale, $\mu\sim 1$~GeV \cite{ref.33}. The value we
take here, $m_s=200$~MeV, is significantly smaller than $\tilde m_s\sim
500$~MeV used in constituent quark models \cite{ref.19, ref.20}. It is
important that the use of current quark masses allows to avoid several
fitting parameters (constituent masses).

\item The value of the string tension $\sigma=0.18$~GeV$^2$ cannot
be used as a fitting parameter, as it is fixed by the slope of
the Regge trajectories for light mesons.

\item The choice of the GE potential is important
for low-lying states. Here we use the vector strong coupling
$\alpha_{\rm B}(r)$ which possesses the asymptotic freedom (AF) property
and freezes at large distances at the value $\alpha_{\rm crit}$.
For higher excitations the choice of $\alpha_{\rm crit}$ becomes
less important; moreover, in many cases the GE potential can be
considered as a perturbation.

\end{enumerate}

In the RSH  $H=H_0+H_{\rm str}$ the unperturbed part
\begin{equation}
 H_0= T(\omega_1, \omega_2)+ V_{\rm B}(r)
\label{eq.02}
\end{equation}
contains the kinetic term $T$ (\ref{eq.01}), where the variables
$\omega_i$ have to be determined from the extremum conditions:
$\frac{\partial H_0}{\partial \omega_i}=0~(i=1,2)$ \cite{ref.28,
ref.34}. Then one finds
\begin{equation}
\omega_i(nL)=\langle\sqrt{ \vep^2+m_i^2}\rangle_{nL} \quad
(i=1,2). \label{eq.03}
\end{equation}
The kinetic energy of a light (strange) quark is denoted as
$\omega_1(nL)=\omega_q(nL)$, and $\omega_2(nL)=\omega_c$ is the
kinetic energy of the $c$ quark; the quantity $\omega_{\rm
red}=\frac{\omega_1\omega_2}{\omega_1+\omega_2},$ and
$\veL=\vel_1+\vel_2$. Then putting $\omega_i$ into
Eq.~(\ref{eq.01}), one arrives at a different form of $T$,
denoted below as $T_{\rm R}$:
\begin{equation}
 T_R=\sqrt{\vep^2+m_q^2} + \sqrt{\vep^2+m_c^2}.
\label{eq.04}
\end{equation}
Rigorously, the expression (\ref{eq.04}) for $T_{\rm R}$ is valid only for
$L=0$, while in general, for $L\neq 0$, $T=T_{\rm R}+T_{\rm str}$ contains the
kinetic energy of the string rotation $T_{\rm str}$. For $L\leq 4$ this term
$T_{\rm str}$ is small compared to $T_{\rm R}$ and can be considered as a
perturbation; its matrix element (m.e.) $\Delta_{\rm str}(nL)=\langle T_{\rm
str}\rangle_{nL}$ is included in the mass formula (\ref{eq.06}). The form $T_{\rm R}$
of the kinetic energy  was suggested in Ref.~\cite{ref.35} and used in many
models \cite{ref.15, ref.36}, while due to our derivation of $T_{\rm R}$ one
can establish the connection between the unperturbed RSH $H_0$ and the RQM,
where the same kinetic term is used.

Then the e.v. $M_0(nL)$ and w.f. are defined by the
spinless Salpeter equation (SSE):
\begin{equation}
 \left[T_{\rm R} + V_{\rm B}(r)\right]\varphi_{nL}=M_0(nl)\varphi_{nL} .
\label{eq.05}
\end{equation}
It is essential that in the RSH approach the spin-averaged meson
mass $M(nL)\equiv M_{\rm cog}(nL)$ is not only defined by the e.v.
$M_0(nl)$ (\ref{eq.05}), but also contains two additional negative
contributions: the string correction $\Delta_{\rm str}(nL)=\langle
H_{\rm str}\rangle_{nL}$ \cite{ref.18, ref.34} and the nonperturbative
self-energy (SE) term $\Delta_{\rm SE}(nL)$ \cite{ref.37}:
\begin{equation}
 M(nl)=M_0(nL) + \Delta_{\rm str}(nL)+ \Delta_{\rm SE}(nL).
\label{eq.06}
\end{equation}
For a given radial quantum number $n$, the string correction
increases for larger $L$, while for a given $L$ it decreases for
higher radial excitations. For the $D(1P)$, $D(1D)$, and $D(1F)$
states their values are equal to $\sim -23$~MeV, $-45$~MeV, and
$-65$~MeV, respectively, which can be obtained using the analytical
expressions for $\Delta_{\rm str}$ from Ref.~\cite{ref.18}. As an
illustration in Table I the masses calculated here for several $D(nL)$
and $D_s(nL)$ states are compared to those from
Ref.~\cite{ref.15}. It appears that differences between them are
mostly due to string corrections, $\sim 40$~MeV,  and our numbers
are closer to the experimental data \cite{ref.2}-\cite{ref.5}.
\begin{table}
\caption{The $D$ and $D_s$ masses for the $1D$, $2P$, and $1F$ states (in
         MeV)\label{tab.1}}
\begin{tabular}{|l|r|r|r|}\hline
   state   &GI\cite{ref.15} &  this paper&   exp. Refs.~[2-5]\\
\hline
   $D(1\,{}^3D_3)$ & 2830  &  2760 &    2762\\
   $D_s(1\,{}^3D_3)$ & 2920 & 2840 & 2860\\
   $D_s(2P_1^{\rm H})$ &  absent  & 3040 & 3044\\
   $D(1\,{}^3F_4)$&3110 & 3030 & absent\\
\hline
\end{tabular}
\end{table}

For the $1P$ and $2P$ states the string corrections are smaller,
$-22$~MeV and $-10$~MeV, respectively, and the masses $M(D_s(2P_1^{\rm
H}))$ calculated here coincide with the experimental mass of
$D_s(3040)$, if this resonance with $J^P=1^+$ is identified as the
higher $2P_1^{\rm H}$ state with $j_s=1/2$ (which has to have a larger
total width), while in the low-mass state with $M(2P_1^{\rm L})=3020$~MeV
the state with $j_s=3/2$ dominates.

Even larger mass differences occur for the yet unobserved states
$D(2\,{}^3P_2)$, and $D(n\,{}^3F_4)~(n=1,2)$, for which the masses
we predict here are $\sim 100$~MeV smaller than in \cite{ref.19} (see
Tables~\ref{tab.2} and \ref{tab.3}).

The perturbative self-energy correction contributes to the current
mass of a heavy quark (it gives $\sim 15\%$ for a $c$ quark
\cite{ref.1}) and moreover there exists a nonperturbative SE
correction to the quark (antiquark) mass. This correction is very
important to provide the linear behavior of the Regge trajectories
\cite{ref.34}.  As shown in Ref.~\cite{ref.37}, this correction is
flavor-dependent and strongly depends on the current quark mass,
being small for a heavy quark and large for a light (strange)
quark \cite{ref.37}:
\begin{equation}
 \Delta_{\rm SE}=-\frac{3\sigma}{2\pi} \left(\frac{\eta_f}{\omega_q(nL)} -
 \frac{\eta_Q}{\omega_Q(nL)} \right).
\label{eq.07}
\end{equation}
The factor $\eta_f$ is determined by the quark current mass and
the vacuum correlation length \cite{ref.37, ref.38}: $\eta_f=1.0$
for a light quark, $\eta_s=0.70$ for the $s$ quark, and
$\eta_c=0.35$ for the $c$ quark. Notice that the number $3/2$
enters the SE term (\ref{eq.07}), instead of the number 2 in
\cite{ref.37}; this change follows from a more exact definition of
the vacuum correlation length \cite{ref.38}.

From Eq.~(\ref{eq.07}) one can see that the kinetic energies
$\omega_i$ play a special role: they determine both the string and
the SE contributions, and also enter all spin-dependent potentials
\cite{ref.39}. In some potential models a negative overall
constant $C_0$ is introduced, which  may play the role of a
self-energy correction, however, such a constant violates the
linear behavior of the Regge trajectories; it is also important
that in the RSH the SE terms decrease for higher states, being
proportional to $\omega_q^{-1}(nL)$.

We use here the ``linear+GE" static potential, $V_{\rm B}(r)$,
which was already tested in a number of our previous works devoted
to heavy-light mesons \cite{ref.18, ref.23} and heavy-quarkonia
\cite{ref.40}:
\begin{equation}
 V_{\rm B}(r)=\sigma r - \frac{4\alpha_{\rm B}(r)}{3 r},
\label{eq.08}
\end{equation}
where the vector coupling $\alpha_{\rm B}(r)$ is taken as in
background perturbation theory \cite{ref.41} with $\alpha_{\rm
crit}=0.50$, which is a bit smaller than $\alpha_{\rm crit}=0.60$
in Ref.~\cite{ref.15}, while a larger value $\alpha_{\rm
crit}=0.84$ was used in Ref.~\cite{ref.19}. In all cases the AF
behavior of the vector coupling is taken into account. Notice that
if a constant value $\alpha_0$ (without AF behavior) is used in
the GE potential, then the value of $\alpha_0$ turns out $\sim
30\%$ smaller than $\alpha_{\rm crit}$.
\section{Higher $D$ mesons}

The masses of higher $D$ excitations are presented in
Tables~\ref{tab.2}, \ref{tab.3} together with results from
\cite{ref.15,ref.19}, where the same $\sigma=0.18$~GeV$^2$ is
used. In these Tables we have omitted results for the ground
states, $1\,{}^1S_0$, $1\,{}^3S_1$, and the $1P$ states, since
they were studied in detail within the same approach in
Ref.~\cite{ref.18}; also for low-lying states their masses do not
differ much in different models, since they are often used as a
fit.

On the contrary, for higher states, large effect takes place when
the constituent quark masses, instead of the current masses, are
used. In Refs.~\cite{ref.15, ref.19} the following masses were
taken:
\begin{eqnarray}
 \mbox{Ref.~\cite{ref.15}}& m_{u,d} =220~{\rm MeV},&
 m_s=419~{\rm MeV}, \quad m_c=1628~{\rm MeV},
\nonumber \\
 \mbox{Ref.~\cite{ref.19}}& m_{u,d} =330~{\rm MeV}, &
 m_s=500~{\rm MeV}, \quad m_c=1550~{\rm MeV},
\nonumber \\
 {\rm this~paper}, & m_{u,d}=0,~\quad\quad\quad &
 m_s=200~{\rm MeV}, \quad m_c=1420~{\rm MeV}.
\label{eq.09}
\end{eqnarray}
Our results are presented in Tables \ref{tab.2} and \ref{tab.3}.
The mass $M(D(1\,{}^3D_3))=2760$~MeV calculated here, coincide
with the experimental mass of $D(2760)$, which is assumed now to
be the $J^P=3^-$ state \cite{ref.13, ref.14}. This value is
smaller than the masses 2863 MeV given in Ref.~\cite{ref.19} and
2830 MeV given in Ref.~\cite{ref.15} and this difference is partly
explained by the string correction, equal to $-45$~MeV.

Much larger differences occur for the excitations with $n=2$ and
$L=2,3$. For example, $M(2\,{}^3D_3)=3212$~MeV is obtained here,
while the value 3335 MeV was predicted in Ref.~\cite{ref.19}, and
this result cannot be explained by a string correction, which is
only $\sim -25$~MeV in this case. From our point of view it
happens due to the use of large constituent mass for a light
quark.

\begin{table}
\caption{The $D$ meson masses $M(nS)~(n=2,3)$ and $M(nD)~(n=1,2)$
(in MeV). The experimental data from \cite{ref.5}; the quark
masses are given in (\ref{eq.09}); $\sigma=0.18$~GeV$^2$ in all
cases.\label{tab.2}}
\begin{tabular}{|l|r|r|r|r|}\hline
state & exp. \cite{ref.5} &  this paper & GI \cite{ref.15} & EFG \cite{ref.19} \\
\hline
 $2\,{}^1S_0$   & 2539   & 2567 & 2580 & 2581\\
 $2\,{}^3S_1$   & 2608   & 2639 & 2640 & 2632 \\
 $3\,{}^1S_0$   & absent & 3065 &      & 3062\\
 $3\,{}^3S_1$   & absent & 3125 &      & 3096\\
 $1\,{}^3D_1$   & ${}^a)$& 2790 & 2820 & 2788 \\
 $1\,{}^3D_3$   & 2763   & 2760 & 2830 & 2863\\
 $1D_2^{\rm H}$ & absent & 2810 &      & 2850\\
 $1D_2^{\rm L}$ & 2750   & 2746 &      & 2806\\
 $2\,{}^3D_1$   & absent & 3215 &      & 3228\\
 $2\,{}^3D_3$   &        & 3212 &      & 3335\\
\hline
\end{tabular}

${}^a)$ The identification of this state is not certain,  because
the quantum numbers of the $D(2760)$ state are not established. It
could be either a $1\,{}^3D_3$ or a $1\,{}^3D_1$ state.
\end{table}

In Tables \ref{tab.2} and \ref{tab.3} we denote by $P_1^{\rm H}$ and
$D_2^{\rm H}$ the high-mass states  with $J=L$, and by $P_1^{\rm L}$ and
$D_2^{\rm L}$  the low-mass states. Each of these states is an admixture
of the state with $j_q=l+1/2$ and $j_q=l-1/2$ in the ${j_q}^2$ scheme,
and for $l=2,3$ the mixing angle between these states appears to be
small (see section~\ref{sect.5}).

For the $2\,{}^3P_2$ state we predict the mass, 2965 MeV, smaller
then the values 3012 MeV in Ref.~\cite{ref.19} and 3035~MeV in
Ref.~\cite{ref.16}. For the states with $L=3$ calculated here, the
mass $M(D(1\,{}^3F_4)=3030$~MeV, is 157 MeV and 80 MeV smaller
than in Refs.~\cite{ref.19} and Ref.~\cite{ref.15}, respectively,
and these large differences can be only  partly explained by the
string correction, equal to $-65$~MeV for the $1F$ states. The
largest difference takes place here for $M(2\,{}^3F_4)=3430$~MeV,
which is much smaller than the value 3610 MeV from
Ref.~\cite{ref.19}. Again, such a large discrepancy cannot be
explained by a string correction, which is $\sim -48$~MeV for the
$2F$ states.

\begin{table}
\caption{The $D$ meson masses $M(nP), M(nF)~(n=1,2)$ (in
           MeV) \label{tab.3}}
\begin{tabular}{|l||r|r|r|}
\hline state &  this paper&  GI  &   EFG   \\\hline
$2\,{}^3P_0$ &  2880  &    &   2919\\
$2P_1^{\rm H}$&      2960 &  &3021\\
$2P_1^{\rm L}$ &  2940&  &2932\\
$2\,{}^3P_2$ & 2965&  &  3012\\
$1\,{}^3F_4$ & 3030 &3110  &3187\\
$1\,{}^3F_2$ &  3088&   & 3090\\
$2\,{}^3F_4$ & 3430&   & 3610\\
\hline
\end{tabular}
\end{table}

In Tables \ref{tab.1}--\ref{tab.4} all FS splittings given are
calculated taking the strong coupling $\alpha_{\rm fs}$ in
the spin-orbit and tensor potentials equal to 0.45.

We can summarize our results for the higher $D$ mesons:
\begin{enumerate}

\item The HF splitting between $D(2\,{}^3S_1)$ and
$D(2\,{}^1S_0)$, equal to 72 MeV, was calculated with the use of the
``universal " strong coupling in the HF potential, $\alpha_{\rm
HF}=0.31$ from Ref.~\cite{ref.27}; this splitting is in full
agreement with experiment.

\item The string corrections, present in the RSH, reduce the
spin-averaged masses of the $nL$ multiplets by -25 MeV, -45 MeV,
-65 MeV for the states with $L=1,2,3$, respectively.

\item Large mass differences for  high excitations like
$D(2\,{}^3D_3)$ and $D(2\,{}^3F_4)$ reach 120 MeV and 180 MeV
compare to the predictions in Ref.~\cite{ref.19}.

\item The recently observed $D(2760)$ and $D(2750)$ resonances are
interpreted as the $D(1\,{}^3D_3)$ and the low-mass $D(1D_2^{\rm L})$
states, where $D(1D_2^{\rm L})$ is in fact the state with $j_q=l+1/2$
(see section~\ref{sect.5}) and therefore should have relatively small
total width, as it is observed in the  BaBar experiments \cite{ref.5}.
It implies that in the $\veL\veS$ scheme the states $D(1\,{}^3D_2)$
and $D(1\,{}^1D_2)$ are mixed with the mixing angle $\theta=40^\circ$.

\end{enumerate}

\section{Higher $D_s$ mesons}

For the $S$-wave states, there are no string corrections,
nevertheless, the mass $M(D_s(3\,{}^1S_0)=3140$~MeV calculated
here, is 79 MeV less than the one given in Ref.~\cite{ref.19} (see
Table \ref{tab.4}). From our point of view, this happens because
of the large constituent mass $\tilde m_s=500$~MeV taken in
Ref.~\cite{ref.19}. To illustrate this effect we have solved the
SSE with two different masses of the $s$ quark: $\tilde
m_s=0.5$~GeV and $m_s=0.2$~GeV, keeping all other parameters the
same. Then the mass difference $\delta(nL)=M_{\rm cog}(nL,m_1=0.5$~GeV$)-
M_{\rm cog}(nL,m_1=0.2$~GeV$)$ appears to
be almost constant for a fixed $L$ and changing $n$:
$\delta(2P)\simeq \delta(3P)=-138$~MeV; $\delta(1D)\simeq
\delta(2D)=-130$~MeV, and $\delta(1F)\simeq \delta(2F)=-120$~MeV,
and thus one may expect mass differences $\sim 100-150$~MeV
to occur between relativistic models with large constituent light
(strange) quark mass compared to the RSH, which uses curent-quark masses.
\begin{table}
\caption{The masses $M(nL)$ (in MeV) for $D_s$ mesons \label{tab.4}}

\begin{tabular}{|l|r|r|r|r|}
\hline
 state &exp.\cite{ref.1}-\cite{ref.4}& this paper&GI \cite{ref.15}
 & EFG \cite{ref.19} \\
\hline
$2\,{}^1S_0$ & 2638$^a$  & 2656  &2670  &   2688\\
$2\,{}^3S_1$ & 2710 & 2728 & 2730 & 2731 \\
             &        2688$^b$ &    &      &       \\
$3\,{}^1S_0$ & absent  &  3140   &    & 3219\\
$3\,{}^3S_1$ & absent  & 3200  &    &   3242  \\
$2\,{}^3P_0$ & absent   &   2970  &     &   3054\\
$2P_1^{\rm H}$ & 3044  &  3040  &   &   3154\\
$2P_1^{\rm L}$ &  absent & 3020   &  & 3067 \\
$2\,{}^3P_2$ & absent &  3045  &   &  3142\\

$1\,{}^3D_1$ &  absent   &  2870  &   2900&  2913\\
$1D_2^{\rm H}$ &  absent & 2885 &        & 2961 \\
$1D_2^{\rm L}$ &  absent & 2828  &    &  2931  \\
$1\,{}^3D_3$ & 2862 & 2840  & 2920&  2973\\
$2\,{}^3D_1$ & absent   &3290 &   & 3383 \\
$2\,{}^3D_3$ & absent   & 3285 &   &   3469\\
$1\,{}^3F_4$ & absent &  3110  &  3190 &    3300 \\
$1\,{}^3F_2$ & absent &3150 &  &  3230\\
$2\,{}^3F_4$ & absent &  3490  &    & 3754\\
\hline
\end{tabular}

\bigskip$^a$ The data of SELEX \cite{ref.26}.
\bigskip$^b$ The data of Belle \cite{ref.3}.
\end{table}

Just for that reason the masses $M(D_s(2\,{}^3D_3)$ and $M(D_s(1\,{}^3F_4))$
are in our calculations $\sim 180$~MeV lower than in
Ref.~\cite{ref.19} and again such a large difference cannot be
explained by the string corrections, which is only $\sim -45$~MeV
for the $D_s(2F)$ state.

Thus one can conclude that a large decrease in the masses of
higher states predicted here, mainly comes from two sources: the string
correction and the use of the current mass for an $s$ quark, which
is significantly smaller than a typical constituent mass $\tilde
m_s\sim 450\pm 50$~MeV.

There exists another characteristic feature of the $D_s$
spectrum~-~for all known states the experimental masses of
$D_s(nL)$ and $D(nL)$ differ by $\sim 100$~MeV. In our
calculations such a spacing $\delta_s(nL)$ comes from two sources;
first, from  different e.v. $M_0(nL)$ of Eq.~(\ref{eq.05})  in
the cases with $m_q=0$ and $m_s=0.20$~GeV, which gives $\sim 50\pm
10$~MeV difference. Secondly, the light and the $s$ quarks have
different nonperturbative SE corrections (negative), which is
$\sim 40\pm 10$~MeV smaller for the $s$ quark as compared to a
light quark. Altogether $\delta_s$ appears to be $\sim 100$~MeV
for low-lying states and smaller, $\delta_s\sim 70-80$~MeV, for
higher states.

Our results about the $D_s$ spectrum can be summarized as follows
\begin{enumerate}

\item The HF splitting between $D_s(2\,{}^3S_1)$ and $D_s(2\,{}^1S_0)$,
calculated with the use of the ``universal" $\alpha_{\rm HF}=0.31$
from Ref.~\cite{ref.27}, gives good agreement with experiment.

\item The resonance $D_s(3044)$ is considered here as the high-mass
state $D_s(2P_1^{\rm H})$, which is dominantly the state with $j_s=1/2$
(see section~\ref{sect.5}) and therefore has to have large total width,
in agreement with experimental value $\Gamma=239\pm 35$~MeV; also its
mass, 3040 MeV, calculated here, is in full agreement with experiment.

\item The resonance $D_s(2860)$ is interpreted as the $D_s(1\,{}^3D_3)$
state and its calculated mass 2840 MeV is in agreement with
experiment. This state with $J^P=3^-$ and $j_s=5/2$ is assumed
to have relatively small total width, as it takes place for
$D_{s2}^*(2573)$. Indeed, the experimental width $\Gamma(D(2860))=48\pm
3$~MeV \cite{ref.2, ref.4} is  small for so high a resonance.

\item The calculated masses of the higher states, like $D_s(2D)$ and
$D_s(2F)$ are 120-200 MeV less than those from Ref.~\cite{ref.19}.

\end{enumerate}

\section{Fine structure splittings}
\label{sect.5}

On a fundamental level, the spin-dependent (SD) potentials
$V_i(r)~(i=1-4)$ have been studied in analytical approaches
\cite{ref.39, ref.42}, and also on the lattice \cite{ref.43},
where the SD potentials are expressed via the vacuum correlators.
When the spin-orbit potential $V_{\rm SO}(r)$ is considered, its
perturbative part can be expressed only through the vector
potential $V_2(r)\equiv V(r)$ if the Gromes relation \cite{ref.44}
is used, and its nonperturbative part is determined by the scalar
confining potential $S(r)=\sigma r$. For the tensor potential the
nonperturbative contribution appears to be very small
\cite{ref.39, ref.43} and it is defined by the perturbative
potential only, usually denoted by $V_3(r)$, which in general case is
not equal to $\left[\frac{V'}{r} -V''\right]$, as it takes place for the 
one-gluon-exchange (OGE) potential (notice, that in the static potential
(\ref{eq.08}) the effective vector coupling $\alpha_{\rm B}(r)$ includes higher order 
perturbative corrections, while these corrections appear to be different for 
different spin-dependent potentials and in OGE approximation they are neglected):
\begin{equation}
 V_3(r)=3 T_0(r)\xi \equiv  \frac{4\alpha_{\rm FS}}{3 r^3}\xi.
\label{eq.10}
\end{equation}
Here the factor $\xi(nL)$ is introduced to show the difference
between $V_3(r)$ and $3T_0(r)$. In heavy quarkonia (HQ) this
factor $\xi$ appears  due to second order perturbative
corrections, being $\xi\simeq 1.30\pm 0.05$, both in charmonium
and bottomonium \cite{ref.30}. However, the value of $\xi$ remains
unknown for heavy-light mesons and the difference between $V_3(r)$
and $3T_0(r)$ may be important for the FS analysis. In HQ for the
$1P$ states the spin-orbit $a_{\rm SO}(1P)$ and the tensor $t(1P)$
m.e. can be extracted from the experimental
masses since all of them are measured with great accuracy.

The study of the FS of the $D(nL)$ and $D_s(nL)$ multiplets is a
more complicated task, because only a few masses are known from
experiment, besides those for the $D(1P)$ and $D_s(1P)$
multiplets. Moreover, many states lie above open thresholds and
may have mass shifts, which change the mass values as compared to
those in single-channel approximation. Nevertheless, a general
analysis of the FS in heavy-light mesons is very useful and allows
to understand better the FS dynamics and make definite conclusions
about mixing angles for the states with $J=L$.

For a multiplet $nL$ the FS is considered here in the basis $j_q^2$,
where the total angular momentum of a light (strange)
quark $j_q$ is diagonal \cite{ref.22}.(Below we use the notation
$j_q\equiv j$). This basis is especially convenient for the
calculation of the mixing angle (denoted as $\phi(nl)$) between
the states with $j=l+1/2$ and $j=l-1/2$, if $J=l$. Another scheme,
$S^2$, is also often used, and in this scheme the notation
$\theta$ for a mixing angle is used here. The relation between
$\theta$ and $\phi$ can be easily established, writing the
high-mass state ($L_J^{\rm H}$) and low-mass state ($L_J^{\rm L}$) with $J=L$
in both schemes. In the $j^2$ basis we write
\begin{eqnarray}
 |J^{\rm H}\rangle & = & \sin\phi |j = l+\frac12\rangle + \cos\phi |j =l-\frac12\rangle,
\nonumber \\
 |J^{\rm L}\rangle & = & \cos\phi |j =l+\frac12\rangle - \sin\phi |j =l-\frac12 \rangle,
\label{eq.11}
\end{eqnarray}
while in the $S^2$ scheme the same physical states are defined as
in Ref.~\cite{ref.12},
\begin{eqnarray}
 |L^{\rm H}\rangle & = & -\sin\theta |^1 L_J\rangle + \cos\theta |^3L_J\rangle,
\nonumber \\
 |L^{\rm L}\rangle & = & \cos\theta |^1 L_J\rangle + \sin\theta |^3L_J\rangle.
\label{eq.12}
\end{eqnarray}
Then taking from Ref.~\cite{ref.22} the relations:
\begin{eqnarray}
 |J=l,j=l-1/2\rangle & = & \sqrt\frac{l+1}{2l+1}|J=l,S=1\rangle
 -\sqrt\frac{l}{2l+1}|J=l, S=0\rangle,
\nonumber \\
 |J=l,j=l+1/2\rangle & = & \sqrt\frac{l}{2l+1}|J=l, S=1\rangle +
 \sqrt\frac{l+1}{2l+1}|J=l, S=0\rangle,
\label{eq.13}
\end{eqnarray}
and inserting them  into Eq.~(\ref{eq.11}), one obtains
\be
 \theta = -\phi + \arccos \sqrt{\frac{l+1}{2l+1}}.
\label{eq.14}
\ee
For $L=1,2,3$ it gives
\be
 \theta(L=1) =-\phi  +35.26^\circ,\quad \theta(L=2) =-\phi + 39.23^\circ,
 \quad \theta(L=3)= -\phi  + 40.89^\circ.
\label{eq.15}
\ee

To determine $\phi$ one needs to know the m.e. of the spin-orbit
and tensor potentials, which are written here in a more general
form than in Ref.~\cite{ref.22}:
\be
 V_{\rm {SO}}=
 \lambda_1(r)2\vel \cdot \ves_1 +\lambda_2(r) 2\vel \cdot \ves_2,
\label{eq.16}
\ee
with
\begin{equation}
 \lambda_1(r)=\frac{1}{4\omega_1^2}\frac{V'-S'}{r}+
 \frac{1}{2\omega_1\omega_2}\frac{V'}{r},\quad
 \lambda_2(r)=\frac{1}{4\omega_2^2}\frac{V'-S'}{r}+
 \frac{1}{2\omega_1\omega_2}\frac{V'}{r}.
\label{eq.17}
\end{equation}
Notice that the kinetic energies $\omega_i$ enter $\lambda_i$ in
Eq.~(\ref{eq.17}) instead of the constituent masses usually used in
potential models. This change follows from the general consideration
of spin-dependent potentials in the RSH \cite{ref.39} and is
important for higher states, decreasing their FS splittings.

For the linear confining potential $S'=\sigma$, while the
perturbative vector potential $V(r)$ is taken in the form,
satisfying the relation  $V'(r)/r=4\alpha_{\rm FS}/r^3\equiv T_0$,
as for the OGE potential, where the vector coupling $\alpha_{\rm FS}$
is considered as an effective coupling. Then the quantity $V_3$ in
the tensor potential,
\begin{equation}
  {V}_{\rm t}(r)= \frac{V_3(r)}{12\omega_1\omega_2}{S}_{12},
\label{eq.18}
\end{equation}
is given in Eq.~(\ref{eq.10}) and the tensor operator is defined
as usual by
\begin{equation}
 {S}_{12}=3\frac{(\vesig_1 \cdot \ver)(\vesig_2 \cdot \ver)}{r^2}
 -\vesig_1 \cdot \vesig_2.
\label{eq.19}
\end{equation}
Later we use for simplicity the notations $\lambda_i(nl)$ for m.e.
$\langle{\lambda}_i(r)\rangle_{nl}$ and
\begin{equation}
 t(nl) = \left\langle\frac{V_3(r)}{3\omega_1\omega_2}\right\rangle_{nl} .
\label{eq.20}
\end{equation}
The spin-orbit m.e. $a_{\rm {SO}}$, given by
\be
 a_{\rm {SO}}(nl)= \lambda_1(nl)+\lambda_2(nl),
\label{eq.21}
\ee
and the tensor m.e. $t(nl)$ fully determine the FS splittings for the states
with $J=l+1$ and $J=l-1$ (in both cases spin $S=1$):
\begin{eqnarray}
 M(J=l+1, S=1) & = & M_{\rm cog} +l a_{\rm SO} - \frac{l}{2 (2l+3)}t,
\nonumber \\
 M(J=l-1, S=1) & = & M_{\rm cog} -(l+1)a_{\rm SO}- \frac{l+1}{2(2l-1)}t.
\label{eq.22}
\end{eqnarray}
For $J=l$ the states with $j=l+1/2$ and $j=l-1/2$ are mixed and
their masses and mixing angles $\phi$ are defined by the matrix
$M_{\rm mix}$:

\begin{equation}
  M_{\rm mix} =
 \left(
 \begin{array}{rr}
 a_{SO}l+\frac{l}{2(2l+1)}[t - 8\lambda_2(l+1)]&
 -(4\lambda_2-t)\frac{\sqrt{l(l+1)}}{2(2l+1)}\\
 -(4\lambda_2-t)\frac{\sqrt{l(l+1)}}{2(2l+1)} \quad&
 -a_{SO}(l+1)+\frac{l+1}{2(2l+1)}(t+8l\lambda_2).
 \end{array}
 \right)
\label{eq.23}
\end{equation}\\
From Eq.~(\ref{eq.23}) one can see that in general the matrix $
M_{\rm mix}$ depends on $a_{\rm SO}$ and $t$, and also on the m.e.
$\lambda_2$, and the value of the factor $4\lambda_2-t$, present
in the off-diagonal m.e., is important for the determination of
the mixing effect.

In the heavy-quark limit there is no mixing, because both $\lambda_2$
and $t$ are going to zero (they are proportional to
$m_Q^{-n}~(n=1,2)$ and may therefore be neglected). Then the high-mass
state H has $j=l+1/2$ while the low-mass state has $j=l-1/2$, if
$a_{\rm SO}$ is positive. However, such a situation with $a_{\rm
{SO}}\gg t$ does not occur even in bottomonium, where for the $1P$
states $a_{\rm SO}(b\bar b, \exp)=13.65\pm 0.39$~MeV coincides
with $t(b\bar b,\exp)=13.13\pm 1.04$~MeV within the experimental
errors, and in charmonium $a_{\rm SO}(c\bar c,\exp)=34.96\pm
0.13$~MeV is even $14\%$ smaller than $t(c\bar c,\exp)=40.63\pm 0.26$~MeV
(their ratio is 0.86).

Such a decrease of the spin-orbit m.e. and the ratio $a_{\rm SO}/t$
occurs due to the negative $-\sigma \langle r^{-1} \rangle_{nL}$ term and
a partial or full cancellation in the m.e.
$\langle\frac{V'-S'}{r} \rangle_{nL}$ is possible.
(Also the m.e. $\langle r^{-3}\rangle_{nL}$, entering the spin-orbit
and tensor m.e., decreases for increasing $n$ and $L$). Therefore it
is of interest to define the quantity
\begin{equation}
 A_{\rm SO}(nL)=\frac43\alpha_{\rm FS}\langle r^{-3} \rangle_{nL}
 - \sigma \langle r^{-1} \rangle_{nL},
\label{eq.24}
\end{equation}
which does not depend on $\omega_i$ and enters $\lambda_i= \frac{A_{\rm
SO}}{4\omega_i^2} + \frac{T_0}{2}~(i=1,2)$. In the $D(D_s)$ mesons the
factor $A_{\rm SO}(nL)$  is negative and its magnitude depends on the
value of $\alpha_{\rm FS}$ taken. Here $\alpha_{\rm FS}=0.45$ is mostly
used, which is a bit larger than $\alpha_{\rm SO}\sim 0.38\pm 0.02$
extracted from the charmonium FS \cite{ref.30}). The values of $A_{\rm
SO}$ can be illustrated by the following numbers:

\begin{enumerate}

\item In bottomonium $A_{\rm SO}\simeq 0.14$~GeV$^3$ is positive
and relatively large, while in charmonium $A_{\rm SO}=\pm
0.01$~GeV$^3$ is already small, even compatible with zero, so that
the ratio $\frac{a_{\rm SO}}{t}=0.86$ is less than unity.

\item For the $D(nL)$ multiplets the factor $A_{\rm SO}$ is always
negative: $\sim -0.017$~GeV$^3$ for the $1P,2P$ states, and
~$\sim - 0.028$~GeV$^3$ for the $1D$ and $1F$ states (for
$\alpha_{\rm FS}=0.45$). This result weakly depends on the quark
masses used.

\item In $ M_{\rm mix}$ a common scale is defined by the tensor
m.e. $t$, and for $t=T_0$ (i.e., $\xi=1.0$) it has values:
$t(1P)=39$~MeV, $t(2P)=29$~MeV, $t(1D)=11.3$~MeV, $t(1F)=5$~MeV.

\item For $L=2,3$ the mixing angle is very small, $|\phi|\leq
1^\circ$, for any reasonable choice of coupling. On the contrary,
for the $nP~(n=1,2)$ states the mixing angle is very sensitive to
$\alpha_{\rm FS}$ used.

\item For small coupling, $\alpha_{\rm FS}\leq 0.30$ the mixing
angle $\phi$ decreases, so that the main uncertainty in any FS
analysis comes from the value of $\alpha_{\rm FS}$ taken, which is not
fixed yet.

\end{enumerate}
In our calculations the following values of the kinetic energies
are obtained for $D(nL)$:
\begin{eqnarray}
 \omega_q(1P) & = & 0.60~~{\rm GeV}, \quad
 \omega_q(1D)  = 0.683~{\rm GeV},\quad \omega_q(1F)=0.757~{\rm GeV},
\nonumber \\
 \omega_c(1P) & = & 1.555~{\rm GeV},\quad \omega_c(1D)=1.588~{\rm GeV}, \quad
 \omega_c(1F)=1.62~~{\rm GeV}.
\label{eq.25}
\end{eqnarray}
For the $D_s$ mesons the FS picture is essentially the same, because
the m.e. for $D_s$, which are important for the FS, coincide within 1-5\%
with those of the $D$ mesons, and therefore the  $D_s$ FS
splittings and mixing angles are practically the same as for the
$D$ mesons (see Tables~\ref{tab.2}-\ref{tab.4}).

We also assume that for a given $nL$ multiplet the masses of the
$M(J=l+1,j=l+1/2)$ and $M(J=l,j=l+1/2)$ states have
no mass shifts (or have small mass shifts), as it happens for
the $D(1P)$ and $D_s(1P)$ multiplets, and therefore the mass
differences between these states,
\begin{eqnarray}
 M(D_{s2}^*(2573))-M(D_{s1}(2535) & = & 37.31\pm 1.0~{\rm MeV},
\nonumber \\
 M(D_2^*(2460) - M(D_1(2422)) & =  & 40.8\pm 1.6~{\rm MeV},
\label{eq.26}
\end{eqnarray}
may be considered as the most stable characteristic of a given
multiplet $nL$; in general this mass difference is
denoted by $\Delta(nl)$:
\begin{equation}
 \Delta(j=l+1/2)=M(J=l+1,j=l+1/2) - M(J=l, j=l+1/2).
\label{eq.27}
\end{equation}
Our calculations show that for the $D(1P)$ states the quantity
$\lambda_1=5.5$~MeV is positive and small, while
$\lambda_2=17.3$~MeV is relatively large, giving  $a_{\rm
SO}(1P)=22.8$~MeV, and $t(1P)=T_0(1P)=38.6$~MeV ($\xi=1.0$), so
that the ratio $a_{\rm SO}/t=0.59$ is smaller than in charmonium,
where this ratio is 0.86.

With the use of $M_{\rm mix}$, Eq.~(\ref{eq.23}), the mass
splittings within the $2P$ multiplet are calculated (see Tables~
\ref{tab.3} and \ref{tab.4}) and the mixing angle depends on
$\alpha_{\rm FS}$, decreasing for smaller coupling: a large angle
$\phi(1P)=-38^\circ$ is obtained for large $\alpha_{\rm FS}=0.60$,
while $\phi(1P)=-12.6^\circ$ for $\alpha_{\rm FS}=0.45$ and
$\phi=-4.2^\circ$ for a smaller $\alpha_{\rm FS}=0.33$.

For higher orbital excitations ($l=2,3,~n=1)$ the nondiagonal
terms in the matrix $ M_{\rm mix}$ appear to be much smaller
than the diagonal m.e. for $\alpha_{\rm FS}=0.45$ and due to this
fact the mixing angles
\begin{equation}
 \phi(1D)=-0.89^\circ, ~~\phi(1F)=-1.0^\circ
\label{eq.28}
\end{equation}
are very small. From Eq.~(\ref{eq.14}) these values of $\phi$
correspond to the following angles $\theta$ between the states
$n\,{}^3L_J$ ($S=1$) and $n\,{}^1L_J$ ($S=0$) with $J=L$:
$\theta(1D)=40^\circ$ and $\theta(1F)=42^\circ$. This result may
be important for the hadronic decays of these resonances
\cite{ref.12}.

The calculated mass differences $\Delta(nL)$, defined in
Eq.~(\ref{eq.27}),
\begin{equation}
 \Delta(1P)=37.6~{\rm MeV},~\Delta(1D)=15~{\rm MeV},~\Delta(1F)=14~{\rm MeV},
\label{eq.29}
\end{equation}
are in good agreement with the experimental numbers:
$\Delta(1P,\exp)=37.6$~MeV \cite{ref.1} and $\Delta(1D,\exp)\simeq
M(D(2760))-M(D(2750))=(11\pm 9)$~MeV \cite{ref.4}.

We do not discuss here the masses of the states with $j=l-1/2$,
which may have large mass shifts. In the single-channel
approximation the mass $M(1P,0^+)$ is 104 MeV smaller than
$M(1P,2^+)$; for $L=2$ almost equal masses $M(1\,{}^3D_3)$ and
$M(1\,{}^3D_1)$ are obtained, while for the $L=3$ states
$M(1\,{}^3F_4)$ is even smaller than $M(\,{}^3F_2)$ (see Tables
\ref{tab.2}, \ref{tab.3}). It does not exclude that because of
possible mass shifts, the physical masses $M(1\,{}^3D_1)$ and
$M(1\,{}^3F_2)$ become smaller than $M(1\,{}^3D_3)$ and
$M(1\,{}^3F_4)$.

In Ref.~\cite{ref.22} for the $1P$ states the approximation
$\lambda_2=\frac{t}{2}$ was used, which in our consideration is
also valid for the $1P$ and $2P$ states. For the $1D$ and $1F$ states the
factor $\lambda_2$ is smaller, $\lambda_2(1D)\sim 0.3t$,
$\lambda_2(1F)\sim 0.2t$, and therefore the factor $4\lambda_2-t$
in the off-diagonal term in Eq.~(\ref{eq.23}) is also smaller.
However, the main reason why a small mixing occurs for $l=2,3$, is
that the diagonal terms appear to be larger than the off-diagonal
terms due to larger algebraic coefficients.

As a result, for $l=2$ and $3$ the high-mass state is dominantly the
state with $j=l-1/2$  and the low-mass state is mostly the state
with $j=l+1/2$.

\section{Conclusions}

The spectra of the $D$ and $D_s$ mesons were studied with the use
of the RSH, where only such fundamental parameters as the string
tension and the quark current masses enter, and the only
uncertainty comes from the freezing constant $\alpha_{\rm crit}$,
which for higher states gives a theoretical error $\la 20$ MeV in
the spin-averaged mass.  We have shown that
\begin{enumerate}

\item The calculated masses of the higher excitations appear to be 50-150
MeV lower than in other RQM with the same string tension
$\sigma=0.18$~GeV$^2$. It occurs for two reasons: first, due to the
string corrections for the states with $L\neq 0$ and secondly,
because we use the current quark masses.

\item Using the $ j^2$ basis, the states with $J=l$
are shown to have very small mixing angles for $l=2$ and $3$: $\phi\approx
-1^\circ$. It means that the states $1\,{}^3L_L$ and $1\,{}^1L_L$
are mixed with $\theta(1D)=40^\circ$ and $\theta(1F)=42^\circ$.

\item The calculated masses of the state $1\,{}^3D_3$ and the
low-mass state $1D_2^{\rm L}$ agree with the new BaBar resonances,
$D(2760)$ and $D(2750)$.

\item The resonance $D_{sJ}^*(2860)$ is considered as the $1D_s(1\,{}^3D_3)$
state and $D_s(3040)$ as the high-mass $2P_1^{\rm H}$ state.

\item For the yet unobserved resonances the following masses are
predicted: $M(D(2\,{}^3P_2))=2965$~MeV, $M(2P_1^{\rm L})=2940$~MeV
$M(D(1\,{}^3F_4))=3030$~MeV and $M(D_s(1\,{}^3F_4))=3110$~MeV.

\end{enumerate}

\begin{acknowledgments}
A.M.B. is grateful to Prof. Yu.A.Simonov for useful discussions.
This work was supported by grant no. NSh-4961.2008.2.
\end{acknowledgments}

\end{document}